\documentclass[useAMS,usenatbib]{mnras}
\usepackage[dvips]{graphicx}
\usepackage[english]{babel}
\usepackage{amsmath}
\usepackage{amssymb}
\usepackage{textcomp}
\usepackage{natbib}
\usepackage{wasysym}

\title[Star formation suppression in merger remnants]{An ALMA view of star formation efficiency suppression in early-type galaxies after gas-rich minor mergers}
\author[F.~van~de~Voort et al.]{Freeke van~de~Voort,$^{1,2,3,4}$\thanks{E-mail: freeke.vandevoort@h-its.org}
Timothy~A.~Davis,$^{5}$ 
Satoki Matsushita,$^{2}$
\newauthor
Kate Rowlands,$^{6}$
Stanislav~S.~Shabala,$^{7}$
James~R.~Allison,$^{8,9}$
Yuan-Sen Ting,$^{10,11,12}$
\newauthor
Anne~E.~Sansom$^{13}$
and Paul~P.~van~der~Werf$^{14}$ \\
$^{1}$Heidelberg Institute for Theoretical Studies, Schloss-Wolfsbrunnenweg 35, 69118, Heidelberg, Germany \\
$^{2}$Academia Sinica Institute of Astronomy and Astrophysics, P.O.\ Box 23-141, Taipei 10617, Taiwan \\
$^{3}$Department of Astronomy and Theoretical Astrophysics Center, University of California, Berkeley, CA 94720-3411, USA \\
$^{4}$Astronomy Department, Yale University, P.O.\ Box 208101, New Haven, CT 06520-8101, USA \\
$^{5}$School of Physics \&\ Astronomy, Cardiff University, Queens Buildings, The Parade, Cardiff, CF24 3AA, UK \\
$^{6}$Department of Physics \& Astronomy, Johns Hopkins University, Bloomberg Center, 3400 N.\ Charles St., Baltimore, MD 21218, \\
\ USA \\
$^{7}$School of Natural Sciences, University of Tasmania, Private Bag 37, Hobart, Tasmania 7001, Australia \\
$^{8}$Sydney Institute for Astronomy, School of Physics A28, The University of Sydney, NSW 2006, Australia \\ 
$^{9}$ARC Centre of Excellence for All Sky Astrophysics in 3 Dimensions (ASTRO 3D) \\ 
$^{10}$Institute for Advanced Study, Princeton, NJ 08540, USA \\
$^{11}$Department of Astrophysical Sciences, Princeton University, Princeton, NJ 08544, USA \\
$^{12}$Observatories of the Carnegie Institution of Washington, 813 Santa Barbara Street, Pasadena, CA 91101, USA \\
$^{13}$Jeremiah Horrocks Institute, University of Central Lancashire, Preston, Lancashire, PR1 2HE, UK \\
$^{14}$Leiden Observatory, Leiden University, P.O.\ Box 9513, NL-2300 RA Leiden, the Netherlands
}

\begin{document}

\date{Accepted 2018 January 24. Received 2018 January 23; in original form 2017 December 21}

\pagerange{\pageref{firstpage}--\pageref{lastpage}} \pubyear{2018}

\maketitle

\label{firstpage}

\begin{abstract}

Gas-rich minor mergers contribute significantly to the gas reservoir of early-type galaxies (ETGs) at low redshift, yet the star formation efficiency (SFE; the star formation rate divided by the molecular gas mass) appears to be strongly suppressed following some of these events, in contrast to the more well-known merger-driven starbursts. We present observations with the Atacama Large Millimeter/submillimeter Array (ALMA) of six ETGs, which have each recently undergone a gas-rich minor merger, as evidenced by their disturbed stellar morphologies. These galaxies were selected because they exhibit extremely low SFEs. We use the resolving power of ALMA to study the morphology and kinematics of the molecular gas. The majority of our galaxies exhibit spatial and kinematical irregularities, such as detached gas clouds, warps, and other asymmetries. These asymmetries support the interpretation that the suppression of the SFE is caused by dynamical effects stabilizing the gas against gravitational collapse. Through kinematic modelling we derive high velocity dispersions and Toomre $Q$ stability parameters for the gas, but caution that such measurements in edge-on galaxies suffer from degeneracies. We estimate merger ages to be about 100~Myr based on the observed disturbances in the gas distribution. Furthermore, we determine that these galaxies lie, on average, two orders of magnitude below the Kennicutt-Schmidt relation for star-forming galaxies as well as below the relation for relaxed ETGs. We discuss potential dynamical processes responsible for this strong suppression of star formation surface density at fixed molecular gas surface density. 

\end{abstract}


\section{Introduction}

In a hierarchical vacuum-dominated cold dark matter ($\Lambda$CDM) universe, mergers play a major role in the assembly of galaxies. Galaxy mergers are capable of inducing strong star formation, fuelling black hole growth, and precipitating morphological transformations, such as the formation of bulges \citep[e.g.][]{Sanders1988, Hernquist1993, Springel2005}. Merger events are often divided into two categories: major mergers with mass ratios larger than $1:4$ and minor mergers with mass ratios below $1:4$. Each of these can be either gas-rich (`wet') or gas-poor (`dry'). Gas-poor mergers are thought to be responsible for both morphological transformation and the build-up of the mass and size of early-type galaxies (ETGs) since $z\approx2$ \citep[e.g.][]{Naab2009, Dokkum2010}. Gas-rich major mergers are usually associated with strong starbursts with extreme star formation rates \citep[SFRs; e.g.][]{Sanders1988, Barnes1991}. 

The star formation efficiency (SFE) is here defined as the amount of star formation per unit \emph{molecular} gas and defined as
\begin{equation}
\mathrm{SFE}=\mathrm{SFR}/M_\mathrm{H_2},
\end{equation}
where $M_\mathrm{H_2}$ is the molecular gas mass. The SFE is the inverse of the gas depletion time, which can be as low as $10^7$~yr in extreme starburst galaxies \citep[e.g.][]{Gao2004}. This unusually efficient star formation is usually explained by the dissipative collapse of the gas to the galaxy centre following a gas-rich major merger. The presence of a pre-existing interstellar medium (ISM) in spiral galaxies causes strong shocks when the gas in the two galaxies collide. These shocks and angular momentum loss drive the gas densities up and result in a nuclear starburst \citep[e.g.][]{Mihos1996}. 

The effect of gas-rich \emph{minor} mergers on the star formation in galaxies has not been quantified as thoroughly. Minor mergers are less violent events, but these too might be expected to increase the SFR and/or SFE \citep[e.g.][]{Mihos1994, Saintonge2012}. \citet{Kaviraj2014} found that minor mergers play an important role in the low-redshift universe, where about 40 per cent of star formation in local spiral galaxies is induced by minor mergers. Such events also dominate the star formation budget in massive early-type systems, where they likely provide the vast majority of the star-forming gas \citep[e.g.][]{Kaviraj2009, Davis2011}.

This work is an extension of that by \citet{Kaviraj2012} and \citet{Kaviraj2013}, who selected a sample of recent minor merger remnants, based on their disturbed optical morphologies. In these recently merged galaxies, the major partner was a gas-poor ETG and the minor partner was a gas-rich dwarf. The median merger mass ratio was estimated to be 1:40, based on gas-to-dust ratios \citep{Davis2015}. The resulting systems are very gas- and dust-rich compared to undisturbed ETGs and have a sizeable molecular gas component ($M_\mathrm{H_2} > 10^9 $~M$_{\astrosun}$). Contrary to expectations, these coalesced systems do not host strong starbursts. The absence of strong merger-induced shocks, due to the gas-poor nature of the original host galaxy, may be responsible for this. More surprising is that the SFE was found to be suppressed by orders of magnitude, with gas depletion times that exceed the Hubble time.

The physical process that is suppressing star formation in these minor merger systems is not yet understood. Clearly such objects do not fit within a framework where the SFE is constant \citep[e.g.][]{Bigiel2011} or one where mergers boost the efficiency of star formation \citep[e.g.][]{Genzel2010, Daddi2010, Saintonge2012}. ETGs have previously been found to have somewhat lower SFE than late-type systems \citep[e.g.][]{Saintonge2012, Davis2014}. \citet{Martig2009} found that this may be caused by gas stability in a spheroidal potential well. This could also be the reason why the star formation rate is less enhanced after minor mergers in galaxies with more prominent bulges \citep{Kaviraj2014}. The recent minor merger systems studied in \citet{Davis2015} and in this work, however, have an order of magnitude greater suppression of the SFE and larger gas fractions than seen in relaxed ETGs. Gas stabilization due to the shape of the potential is therefore unlikely to be the reason for the low SFEs, at least in these extreme objects. 

Star formation is a dynamical process. The cold, dense gas in galaxies follows a turbulent cascade down to the smallest scales, where it fragments and eventually forms stars. Various dynamical and environmental processes can work to suppress this cascade and affect the resulting SFE. \citet{Davis2015} suggested that these objects may have been caught at a very specific phase in their evolution, when gas is still free-streaming towards the galaxy centre, but has not yet settled. Such streaming motions have been shown to suppress the efficiency of star formation \citep[e.g.][]{Meidt2013}. In this scenario, the gas-free nature of the early-type progenitor explains why star formation suppression after minor mergers has not been observed in other systems. 

This explanation is not unique, however. Shear induced by rotation can prevent the gravitational collapse of gas clouds or pull them apart, suppressing star formation \citep{Toomre1964, Seigar2005}. Additionally, gravitational heating, i.e.\ conversion of potential energy to thermal energy during a merger, can deposit energy into the gas via weak shocks \citep[e.g.][]{Johansson2009}. Such energy could stall gravitational collapse and suppress star formation. Alternatively, although no evidence for strong nuclear activity has been found in these objects \citep{Shabala2012}, active galactic nuclei (AGN) are also expected to be triggered by mergers \citep{Kaviraj2015}. Therefore, nascent feedback from a central black hole may be acting to suppress star formation. Determining which, if any, of these processes are acting to suppress star formation in ETGs after a minor merger will allow us to understand its importance for merger-driven star formation in all types of galaxies.

In this paper we present results based on new observations with the Atacama Large Millimeter/submillimeter Array (ALMA) of six minor merger remnants from \citet{Davis2015}. We selected those objects with the most suppressed SFEs (below 10$^{-10}$ yr$^{-1}$). Their depletion times are above 10 Gyr, longer than those of `normal' ETGs, such as those in the ATLAS$^\mathrm{3D}$ sample \citep{Davis2014}, by a factor of three. They have typical stellar masses of about 10$^{11}$ M$_{\odot}$, with molecular gas fractions between 1~and 25~per cent.

In Section~\ref{sec:obs} we describe the targeted galaxies and the ALMA observations. In Section~\ref{sec:results} we show the resolved ISM morphologies and velocity structures, which often exhibit clear disturbances, and present results from our kinematic modelling. We discuss which process is likely responsible for the low SFEs and conclude in Section~\ref{sec:concl}. In this work, we assume a cosmology with $H_0=71$~km~s$^{-1}$~Mpc$^{-1}$, $\Omega_m=0.27$ and $\Omega_\Lambda=0.73$, consistent with previous work.

\section{Observations} \label{sec:obs}

\subsection{Sample}

\begin{figure*}
\center
\includegraphics[scale=0.6]{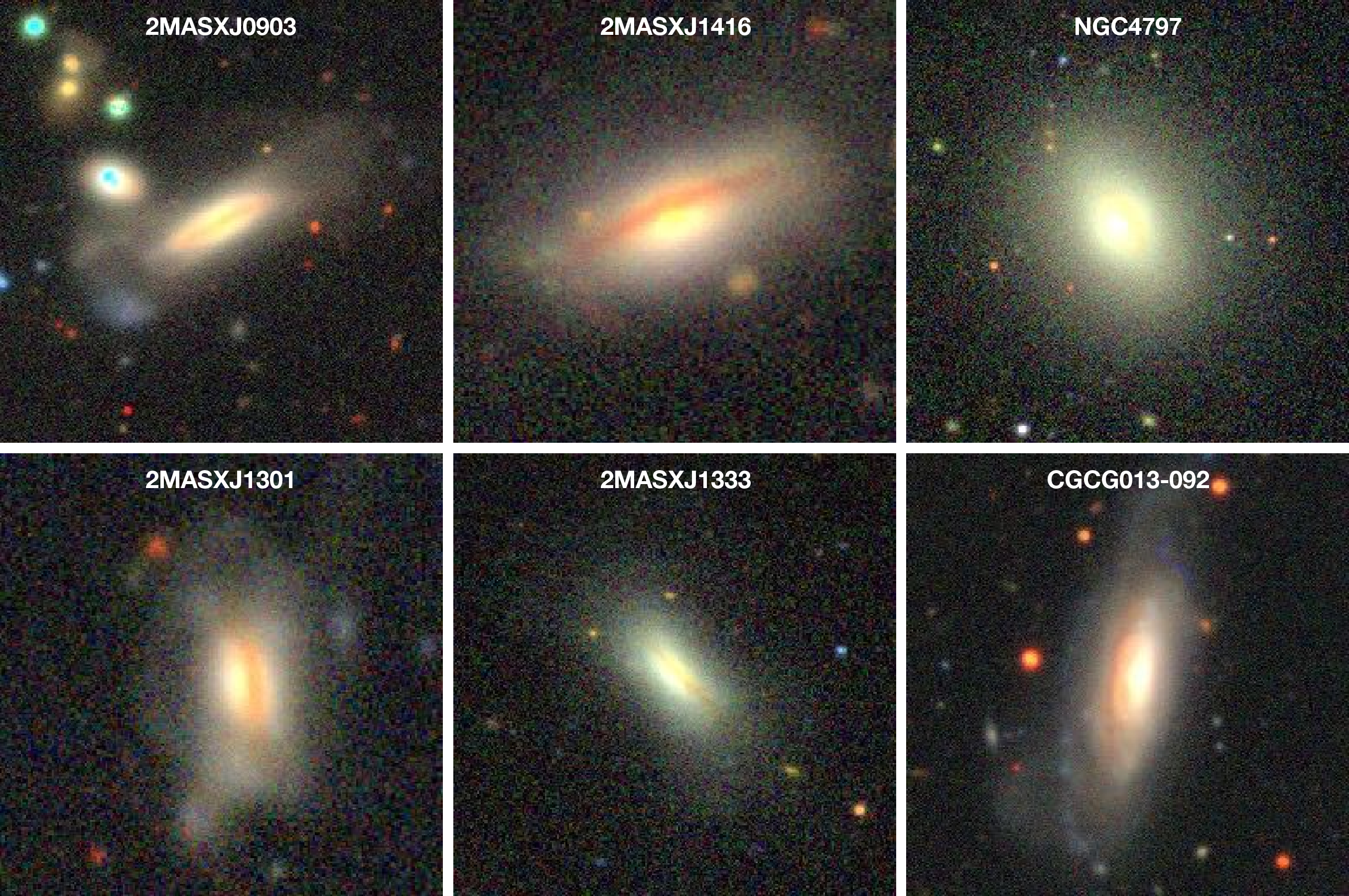}
\caption {\label{fig:3colour} Three-colour imaging of our target sources at fixed spatial scale of $66\times66$~kpc$^2$ from DECaLS ($grz$), when available, and from SDSS ($gri$) for galaxies NGC4797 and 2MASXJ1333. Each panel is labelled with (a shortened version of) the source name. The dust lanes are clearly visible for all galaxies and are almost edge-on. Disturbances in the stellar light are also visible in certain panels, providing evidence for a recent minor merger.}
\end{figure*} 
The galaxies targeted in this study were originally selected from the Sloan Digital Sky Survey (SDSS) photometry by \citet{Kaviraj2012} as part of the Galaxy Zoo project \citep{Lintott2008}. The objects are massive, bulge-dominated galaxies with large dust lanes obscuring part of the optical light. The parent sample selected the most dust-rich objects using 250~$\mu$m luminosities observed with the \textit{Herschel Space Telescope} \citep{Kaviraj2013}. From the galaxies observable from ALMA's location, we selected the six galaxies with the lowest SFEs based on previous Institut de Radioastronomie Millim\'{e}trique (IRAM) 30~m data \citep{Davis2015}. Three-colour ($66\times66$~kpc$^2$) images of our six targets are shown in Figure~\ref{fig:3colour}, using the Dark Energy Camera Legacy Survey (DECaLS; \citealt{DECaLS2016}; in bands $grz$), when available, and from SDSS (\citealt{SDSSDR72009}; in bands $gri$) for galaxies NGC4797 and 2MASXJ1333. Besides the clear dust lanes, indicating the likely presence of edge-on gas discs, some of the images also reveal clear disturbances of the stellar light in the outskirts of the galaxies, which point to recent minor merger events \citep[see][]{Kaviraj2012}.

\begin{table*}
\begin{center}
\caption{\label{tab:gal} \small Properties of our selected galaxies: source identifier, redshift, angular diameter distance, stellar mass, SFR, SFE, and conversion factor from arcsec to kpc. The errors given are $1\sigma$ errors on the measurements.}
\vspace{-3mm}
\begin{tabular}[t]{lllllccr}
\hline
\hline \\[-3mm]
source identifier & $z$ & $D_L$ & $M_\mathrm{star}$ & $M_\mathrm{H_2}$ & SFR & SFE & conversion \\
& & (Mpc) & (M$_{\astrosun}$) & (M$_{\astrosun}$) & (M$_{\astrosun}$~yr$^{-1}$) & (yr$^{-1}$) & (kpc/\arcsec) \\
\hline \\[-4mm]                                                                                                                                       
2MASXJ09033081--0106127 & 0.040 & 161 & $10^{10.9\pm0.1}$ & $10^{9.64\pm0.07}$ & $0.09^{+0.05}_{-0.01}$ & $(2.06^{+1.19}_{-0.40})\times10^{-11}$ & 0.78 \\[1mm] 
2MASXJ14161186+0152048  & 0.082 & 315 & $10^{11.2\pm0.1}$ & $10^{9.89\pm0.11}$ & $0.02^{+0.06}_{-0.01}$ & $(2.58^{+7.77}_{-1.45})\times10^{-12}$ & 1.53 \\[1mm] 
NGC4797                               & 0.026 & 106 & $10^{11.2\pm0.1}$ & $10^{9.17\pm0.08}$ & $0.03^{+0.03}_{-0.01}$ & $(2.03^{+2.06}_{-0.77})\times10^{-11}$ & 0.52 \\[1mm] 
2MASXJ13010083+2701312  & 0.078 & 301 & $10^{10.9\pm0.1}$ & $10^{10.31\pm0.05}$ & $0.44^{+0.44}_{-0.37}$ & $(2.16^{+2.17}_{-1.83})\times10^{-11}$ & 1.46 \\[1mm] 
2MASXJ13333299+2616190  & 0.037 & 150 & $10^{10.8\pm0.1}$ & $10^{9.71\pm0.05}$ & $0.01^{+0.06}_{-0.01}$ & $(1.95^{+11.70}_{-1.95})\times10^{-12}$ & 0.73 \\[1mm] 
CGCG013--092                     & 0.035 & 142 & $10^{10.9\pm0.1}$ & $10^{9.84\pm0.04}$ & $0.28^{+0.06}_{-0.04}$ & $(4.05^{+0.94}_{-0.69})\times10^{-11}$ & 0.69 \\[0mm] 
\hline
\hline                                                                                                                                                
\end{tabular}
\end{center}
\end{table*}   
The properties of our target galaxies are taken from \citet{Davis2015} and presented in Table~\ref{tab:gal}. The source name, redshift, luminosity distance, stellar mass, SFR, SFE, and conversion factor from arcsec to kpc are listed for each source. The SFRs are derived by modelling the spectral energy distribution from the ultraviolet to the far-infrared with energy-balance code \textsc{magphys} \citep{Cunha2008} and therefore typically sensitive to timescales of approximately 100~Myr \citep{Kennicutt2012}. The errors on the SFE are unfortunately fairly large and the SFEs of the six galaxies are consistent with each other within $2\sigma$. We therefore do not expect to find strong trends with SFE within our sample.
 
\subsection{ALMA data}

We observed the $^{12}$CO(1--0) line in our six dust lane ETGs with ALMA, as part of programme 2015.1.00320.S. ALMA's 12~m antennas were used in a compact configuration, resulting in an approximately 1\arcsec\ beam and sensitivity to emission on scales up to $\approx30-40\arcsec$. A 1850 MHz correlator window was placed over the CO(1--0) line, yielding a continuous velocity coverage of about $1600$~km~s$^{-1}$ with a raw velocity resolution of $\approx1.0$~km s$^{-1}$. This is sufficient to properly cover and sample the line. Three additional 2~GHz wide low-resolution correlator windows were simultaneously used to potentially detect continuum emission (see Section~\ref{sec:cont}).

The raw ALMA data were calibrated using the standard ALMA pipeline, provided by the ALMA regional centre staff. We then used the Common Astronomy Software Applications (\textsc{casa}) package to image the resulting visibility files for each track, producing a three-dimensional RA-Dec-velocity data cube (with velocities determined with respect to the rest frequency of the $^{12}$CO(1--0) line). In this work we bin the raw data to a channel width of 10~km~s$^{-1}$ and use pixels of 0.3\arcsec\ (giving us approximately 4 pixels across the synthesized beam). 

The data presented here were produced using Briggs weighting with a robust parameter of 0.5. We attempted to detect continuum emission from these sources over the full line-free bandwidth. If present, we subtracted the continuum from the line data in the $uv$ plane using the \textsc{casa} task \textsc{uvcontsub}. The continuum-subtracted dirty cubes were cleaned in regions of source emission (identified interactively) to a threshold equal to the root-mean-square (RMS) noise of the dirty channels. The clean components were then re-convolved using a Gaussian beam of full-width at half-maximum (FWHM) equal to that of the dirty beam. Finally, the residuals were added back into the clean components. This produced the final, reduced, and fully calibrated $^{12}$CO(1--0) data cubes.

The molecular masses used here assume a Galactic $X_\mathrm{CO}$ factor, which quantifies the conversion between CO integrated line flux and $\mathrm{H}_2$ mass. However, $X_\mathrm{CO}$ has been shown to vary with metallicity (see \citealt{Bolatto2013} for a review). We unfortunately do not have an independent measure of the gas-phase metallicity in our galaxies, which may be sub-solar. We could therefore be underestimating the amount of molecular gas present in these systems, possibly by an order of magnitude \citep{Leroy2007}. This would only reduce the SFEs even further and make our targets even more extreme objects.

\subsection{Line detections}

\begin{table*}
\begin{center}
\caption{\label{tab:obs} \small Parameters for our ALMA observations: source identifier, date of observation, total observation time, amplitude calibrator, bandpass calibrator, phase calibrator, beam size, and RMS noise in 10~km~s$^{-1}$ channels of each source.}
\vspace{-3mm}
\begin{tabular}[t]{llllllcc}
\hline
\hline \\[-3mm]
source identifier & obs.\ date & $t_\mathrm{obs}$ & amplitude & bandpass & phase & $\theta_\mathrm{beam}$ & RMS \\
& & (mins) & calibrator & calibrator & calibrator & (\arcsec) & (mJy beam$^{-1}$) \\
\hline \\[-4mm]                                                                                                                                       
2MASXJ09033081--0106127 & 2016-03-19 & 29.36 & J0750+1231 & J0854+2006 & J0909+0121 & $1.91\times1.32$ & 0.9 \\
2MASXJ14161186+0152048  & 2016-04-09 & 47.35 &  Ganymede   & J1337-1257 & J1410+0203  & $1.78\times1.46$ & 0.8 \\
NGC4797                               & 2016-03-21 & 62.73 & Callisto        & J1229+0203 &  J1303+2433  & $1.80\times1.35$ & 1.0 \\
2MASXJ13010083+2701312  & 2016-05-04 & 29.86 & Callisto & J1229+0203 &  J1303+2433  & $2.05\times1.29$ & 1.1 \\
2MASXJ13333299+2616190  & 2016-05-04  & 27.15 &   J1256-0547 & J1229+0203 & J1333+2725   & $2.35\times1.03$ & 1.5 \\
CGCG013--092                     & 2016-03-20 & 34.34 &  Callisto & J1220+0203 & J1229+0203  & $1.62\times1.31$ & 1.0 \\[-1mm]
\hline
\hline                                                                                                                                                
\end{tabular}
\end{center}
\end{table*}   
Table \ref{tab:obs} summarizes our observational parameters. For each source, we list its full name (taken from the NASA/IPAC Extragalactic Database\footnote{https://ned.ipac.caltech.edu}), the dates the galaxies were observed, the total integration time, the calibrator sources used (for the amplitude, bandpass, and phase), the beam size, and the RMS noise reached in 10~km~s$^{-1}$ channels. The data quality and precision are sufficient to resolve the spatial structure of the molecular ISM and its line-of-sight velocity in all of our target galaxies using the CO(1-0) line. No other lines were detected. We use these data to detect and quantify morphological and kinematic disturbances in the gas in Section~\ref{sec:results}.

\subsection{Continuum detections and upper limits} \label{sec:cont}

\begin{table}
\begin{center}
\caption{\label{tab:dust} \small Continuum detections or upper limits: source identifier, observed frequency, and 3~mm continuum flux (or $3\sigma$ upper limit) of each source. In case of a detection, we analyze the upper and lower sideband (USB and LSB) seperately.}
\vspace{-3mm}
\begin{tabular}[t]{lll}
\hline
\hline \\[-3mm]
source identifier & observed $\nu$ & continuum \\
& (GHz) & ($\mu$Jy) \\
\hline \\[-4mm]                                                                                                                                       
2MASXJ09033081--0106127 & 97.9 (LSB) & $332\pm39$ \\
                                              & 110.0 (USB) & $370\pm50$ \\
2MASXJ14161186+0152048  & 99.7 & $<22$ \\
NGC4797                               & 99.3 (LSB) & $218\pm17$ \\
                                              & 111.5 (USB) &$295\pm29$ \\
2MASXJ13010083+2701312  & 100.1 & $<23$ \\
2MASXJ13333299+2616190  & 104.2 & $<34$ \\
CGCG013--092                     & 104.4 & $<80$ \\[-1mm]
\hline
\hline                                                                                                                                                
\end{tabular}
\end{center}
\end{table}   
Although the focus of this work is on the molecular gas, we attempted to detect continuum emission as well. Table \ref{tab:dust} summarizes our findings and lists, for each source, our measurement or upper limit of the 3~mm continuum flux (right column) and the frequency at which this was observed (middle column). In case of a detection, we list the flux from the upper and lower side band (USB and LSB) separately, but combined them for our upper limits. We detected the continuum for galaxies 2MASXJ0903 and CGCG013--092 ($\approx300$~$\mu$Jy), which are both point sources at 3~mm. By comparing to the far-infrared fluxes from \citet{Rowlands2012}, we interpret this emission as arising from the Rayleigh-Jeans tail of the dust emission. We do not discuss this further here, because this is beyond the scope of this work. The lack of synchrotron-dominated spectra at 3~mm (non-detections or a spectral slope inconsistent with synchrotron emission) argues against the presence of strong AGN, but we cannot exclude low-luminosity AGN.

\section{Results} \label{sec:results}

\begin{figure*} 
\center
\includegraphics[scale=0.605]{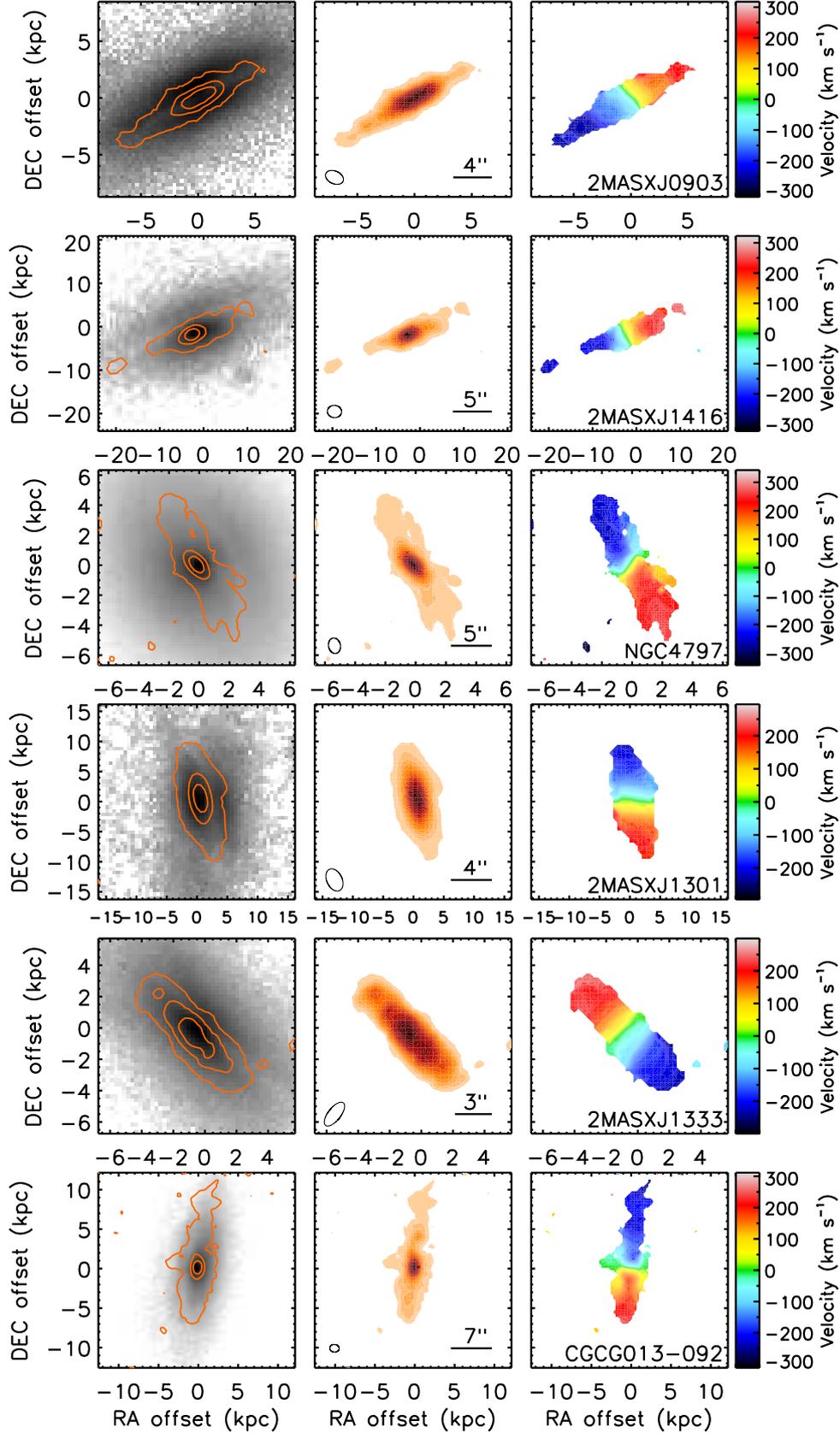}\\
\caption{\label{fig:img} Images of the CO(1-0) data for all six galaxies. Left panels: SDSS $r$-band image overlaid with ALMA CO(1-0) contours. Middle panels: CO(1-0) flux contours. Right panels: Line-of-sight velocity contours of the gas. The contours displayed range from 3 times the RMS level to the maximum value in 3 (left) or 19 (middle) linearly spaced increments. Most of the galaxies exhibit morphological disturbances, such as detached gas clouds, warps, and other asymmetries, indicating that the gas is not completely settled.}
\end{figure*}
The aim of this work is to resolve the spatial and kinematic structure of the molecular gas in the selected dust lane ETGs in order to narrow down the cause of the suppression in their SFEs. Figure~\ref{fig:img} therefore shows the new ALMA observations of the six selected galaxies. Left and middle panels show CO(1-0) flux contours in orange, in the left panels combined with the SDSS $r$-band image in greyscale. The contours displayed range from 3 times the RMS level to the maximum value in 3 (left) or 19 (middle) linearly spaced increments. The right panels show the CO(1-0) line-of-sight velocity contours centred on the kinematic centre as determined from their kinematic modelling (see Section~\ref{sec:kin}). 

\begin{figure*} 
\center
\includegraphics[scale=0.8]{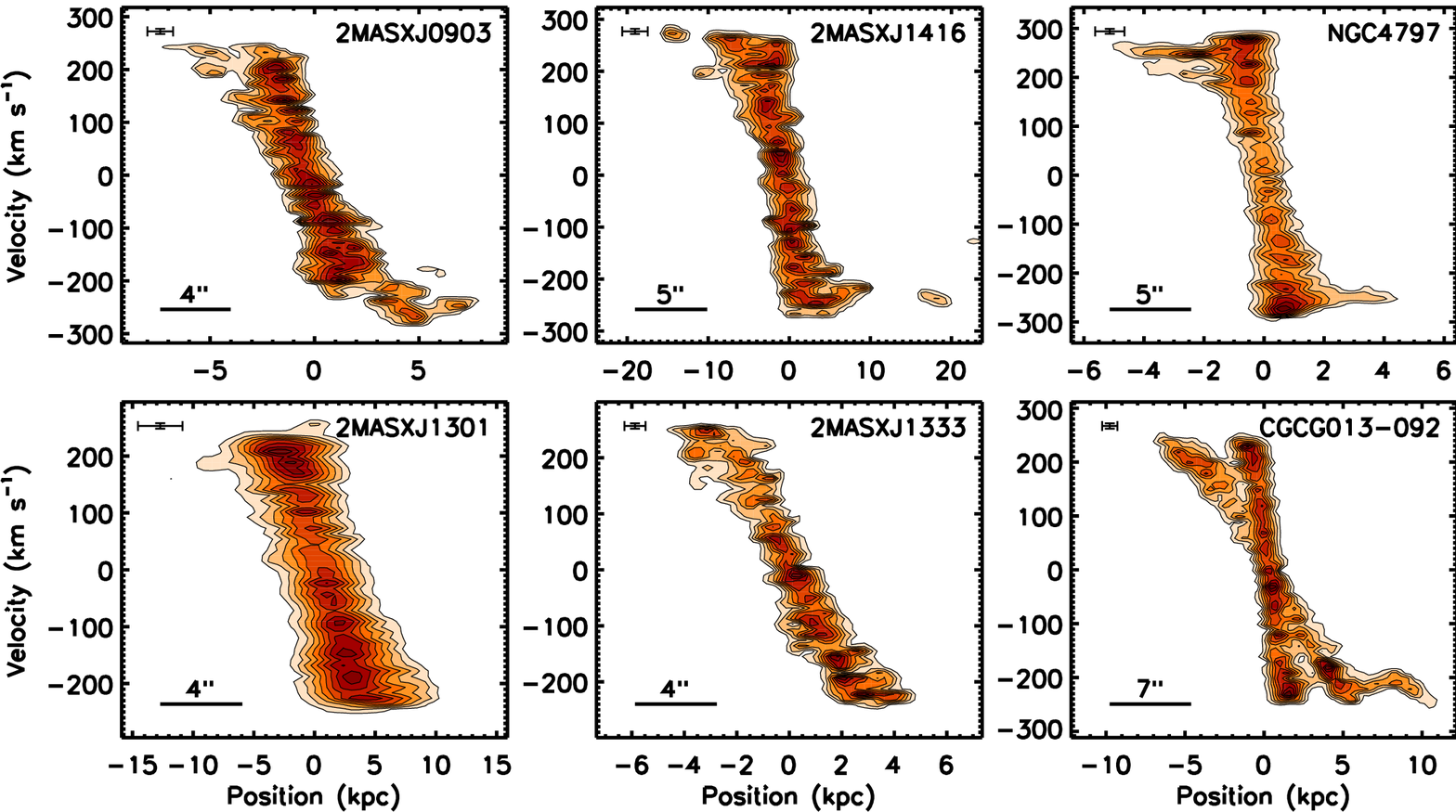}\\
\caption{\label{fig:PVD} Position-velocity diagram for our selected galaxies extracted along the major axis. Contour levels are evenly spaced between 3 times the RMS level and the peak flux in 9 increments. The majority of the observed galaxies show asymmetries and disturbed features with velocities likely offset from the circular velocity of the galaxy. 2MASXJ1301 is fairly regular, but has detectable flux out to 10 kpc on one side, whereas it only extends to 8 kpc on the other side. 2MASXJ1333 is the only galaxy that shows no sign of kinematic disturbances in CO(1-0) at this resolution. This shows that, for most of our dust lane ETGs, not all the molecular gas is dynamically relaxed. This argues in favour of dynamical stabilization as a way of suppressing the SFE in recent minor merger remnants.}
\end{figure*}

As expected, the gas is coincident with the dust lane seen in absorption against the stellar light, which means that both likely have the same origin. The images show that the  majority of the galaxies' molecular gas discs are asymmetrical. 2MASXJ1416 has detached gas clouds, the low-surface brightness emission extends further from the centre on one side of the disc in 2MASXJ0903 and CGCG013--092, and NGC4797 and 2MASXJ1301 exhibit a small warp in their centres (best seen by the tilt in the zero velocity contour in the velocity maps). Only galaxy 2MASXJ1333 shows no sign of asymmetries. The disturbed kinematics can also be identified in the position-velocity diagrams (see Figure~\ref{fig:PVD}). This provides visual evidence that the molecular gas is not completely relaxed. It also supports our conclusion that the gas was brought in by a gas-rich minor merger in the recent past. 

It is interesting to compare the extent of the molecular gas in our sample to that of relaxed ETGs in the ATLAS$^\mathrm{3D}$ sample. 
We measure the radius of the relaxed gas disc by fitting a kinematic model to the data cube, which only reproduces the symmetric, undisturbed part (see Section~\ref{sec:kin}). The radius of the undisturbed gas is defined as the radius at which the model surface brightness falls below 10~M$_\odot$pc$^{-2}$, which is the surface brightness limit used by \citet{Davis2013}. The total extent, including the morphologically and kinematically disturbed part, is measured directly from the data, also using the surface brightness limit of 10~M$_\odot$pc$^{-2}$. 
For the relaxed galaxies, the radius of the molecular gas disc normalized by the radius of the 25~mag~arcsec$^{−2}$ isophote in the $B$ band is $R_\mathrm{CO}/R_{25}=0.16$ \citep{Davis2013}. Unsurprisingly, the gas discs in our selected dust lane ETGs are more extended. For the undisturbed part of our gas discs, we find an average $R_\mathrm{CO}/R_{25}=0.28$ and if we include the disturbed gas this increases to $R_\mathrm{CO}/R_{25}=0.53$. This shows that our sample galaxies are extreme, not just in their total molecular gas mass, but also in the extent of their molecular gas.

\begin{figure*} 
\center
\includegraphics[scale=0.8]{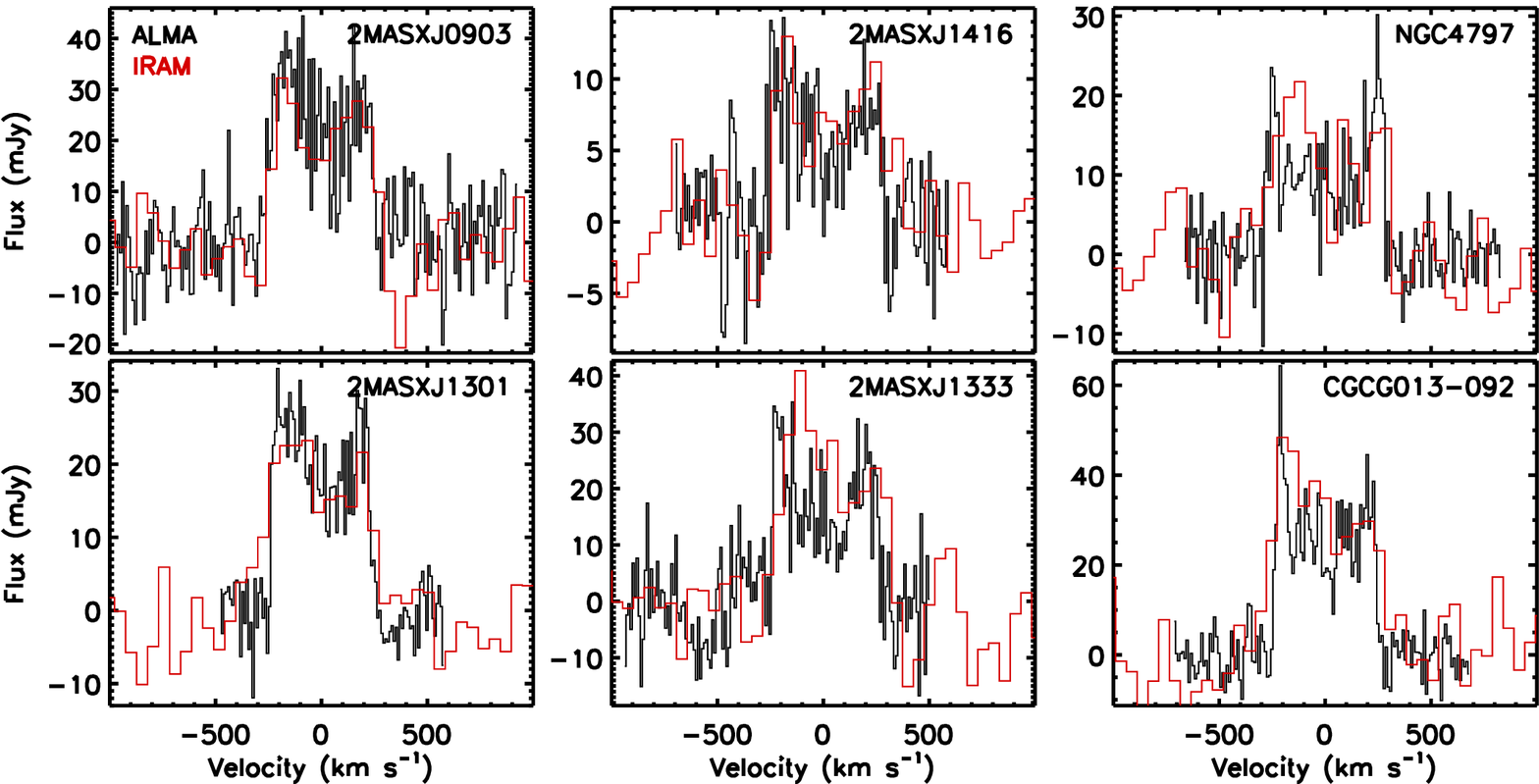}\\
\caption{\label{fig:spec} Spectra of the CO(1-0) data for all six galaxies with our new ALMA data shown in black and previously obtained (lower resolution) IRAM data in red. The flux detected in our new interferometric observations is similar to that from the previous single dish measurements. This means that we are likely capturing molecular gas structures on all important scales with ALMA.}
\end{figure*}
In Figure~\ref{fig:spec} we compare our ALMA CO(1-0) flux measurements with those previously obtained with the IRAM 30~m telescope \citep{Davis2015}, showing the CO(1-0) flux as a function of velocity, where the centre is obtained from our kinematic modelling (see Section~\ref{sec:kin}). The ALMA spectra were made by extracting a rectangular region based on the images shown in Figure~\ref{fig:img}, then summing the unclipped data cube spatially, and dividing by the beam area. Both observations show the typical double-horned spectrum of rotating galaxies. Our interferometric observations are similar to the single dish measurements. The total flux is consistent within uncertainties, which indicates that we are likely not resolving out any structure on large scales. The spectra of galaxies 2MASXJ0903, 2MASXJ1416, 2MASXJ1301, and CGCG013--092 have an asymmetric profile, reflecting their asymmetric morphologies.

To study the rotation curve of the molecular gas, Figure~\ref{fig:PVD} shows the major-axis position-velocity diagram of each of our target galaxies. This was created by extracting a slice from the cube at the estimated position angle, with a width of 5 pixels, which were then summed spatially. Contour levels are evenly spaced between 3 times the RMS level and the peak flux in 9 increments. The disturbed structures, previously identified along the major axis in the images in Figure~\ref{fig:img}, can also be seen here. According to our best-fit kinematic model (see Section~\ref{sec:kin}), they do not follow the general galaxy rotation curve, although accurate stellar photometry is needed to model the stellar mass and be absolutely certain about this. 4 out of our 6 targets clearly show asymmetrical features at large radii -- galaxies 2MASXJ0903, 2MASXJ1416, NGC4797, and CGCG013--092. The asymmetry of galaxy 2MASXJ1301 is more subtle, but can be identified by noticing that the emission extends from $-8$~kpc to $+10$~kpc. Galaxy 2MASXJ1333 is the only object for which, at the current resolution, all the detected molecular gas seems to be part of a regular, relaxed gas disc. 

A large fraction of the molecular gas is located in the rising part of the rotation curve. This is consistent with samples of relaxed ETGs (from ATLAS$^\mathrm{3D}$ and EDGE-CALIFA), which show that the SFE decreases as this fraction increases \citep{Davis2014, Colombo2017}. This could suggest that a dynamical effect, such as the higher shear rate in this region, acts to stabilize the gas against collapse. However, the dust lane ETGs have SFEs that are much lower than those in ATLAS$^\mathrm{3D}$, so this is likely not the dominant mechanism for our extreme SFE suppressed galaxies.

The molecular gas which exists outside the central, undisturbed part of the discs appears not to be in dynamical equilibrium. Instead, it is likely flowing towards or away from the centre. Because the disturbed structures are oriented along the major axis, we consider the most likely scenario that this gas is accreting.

\subsection{Kinematic modelling} \label{sec:kin}

To learn more about the dynamical state of the gas and its physical parameters, we aim to construct kinematic models that reproduce our CO observations. In order to model the kinematics of the gas, we use a forward modelling approach. We assume that the gas is in a circularly symmetric disc and therefore do not try to model the asymmetrical features seen in Figure~\ref{fig:img} and~\ref{fig:PVD}. Because the disturbances are low surface density features, they do not strongly affect the model, which aims to reproduce the majority of the molecular gas (i.e.\ the central disc). 

We used the KINematic Molecular Simulation (KinMS\footnote{https://github.com/TimothyADavis/KinMS}) tool \citep{Davis2013}, with which we produce mock observations of a theoretical gas distribution with the same beam, pixel size, and velocity resolution as our ALMA observations. We then compare the mock observations to the real data and explore the parameter range with the Markov Chain Monte Carlo (MCMC) code KinMS\_MCMC that couples to the KinMS routines in order to get the full Bayesian posterior probability distribution. This code fits the entire CO data cube produced by ALMA, rather than simply the position-velocity diagram shown in Figure~\ref{fig:PVD}.

The mass budget in these ETGs is dominated by their stars. Due to the lack of sufficiently high resolution optical imaging, we model a galaxy's rotation curve by assuming it follows an arctangent \citep[e.g.][]{Swinbank2012}. We force the curve to reach its maximum faster by imposing
\begin{equation}
\begin{split}
v &= v_\mathrm{circ, flat} \mathrm{tan}^{-1}\Big(\frac{R}{r_\mathrm{norm}}\Big) \ \ \ \mathrm{for} \ \mathrm{tan}^{-1}\Big(\frac{R}{r_\mathrm{norm}}\Big)  < 1 \\
 &= v_\mathrm{circ, flat} \ \ \ \ \ \ \ \ \ \ \ \ \ \ \ \ \ \ \ \ \ \  \mathrm{for} \ \mathrm{tan}^{-1}\Big(\frac{R}{r_\mathrm{norm}}\Big)  \ge 1,
\end{split}
\end{equation}
where $v_\mathrm{circ, flat}$ is the asymptotic (or maximum) circular velocity in the flat part of the rotation curve, $R$ is the radius, and $r_\mathrm{norm}$ is the normalization radius. We fit $v_\mathrm{circ, flat}$ and $r_\mathrm{norm}$ to best match the gas velocities. We repeated our kinematic modelling based on available long-wavelength optical or near-infrared image (from SDSS or \textit{Spitzer}) and found best fit values similar to those using the arctangent model and presented in Table~\ref{tab:Q}.

We furthermore fit the total flux, kinematic centre, systemic velocity, position angle, and inclination of the gas disc. We additionally fit a model of the gas surface brightness profile and gas velocity dispersion (assumed to be constant throughout the disc), in order to reproduce the bulk properties of the disc. For the galaxies 2MASXJ0903, 2MASXJ1416, NGC4797, and 2MASXJ1301, we find a reasonable fit by using an exponential gas surface brightness profile. This simple form has previously been shown to be appropriate in most ETGs \citep{Davis2013} and provides a good match to the observed morphology and velocity of the gas in these objects. For galaxy 2MASXJ1333, however, we find that we additionally need an outer surface density cutoff in order to fit the data. Object CGCG013--092 clearly shows an X-shape in its position-velocity diagram, which is typical for molecular gas within a bar. We model this with a central ring and an outer exponential profile. 

We use the derived centre, systemic velocity, and position angle to centre the previous figures. With the derived velocity dispersion and gas profile, we try to estimate whether or not the observed gas discs are stable against gravitational collapse. In order to do this, we calculate the Toomre $Q$ parameter \citep{Toomre1964} by evaluating
\begin{equation}
Q=\dfrac{\kappa \sigma_\mathrm{gas}}{\pi G \Sigma_\mathrm{gas}}.
\end{equation}
Here, $G$ is the gravitational constant, $\Sigma_\mathrm{gas}$ is the surface density of the molecular gas, and $\kappa$ is the epicyclic frequency (i.e.\ the frequency at which a gas parcel will oscillate when radially displaced). The latter is calculated by
\begin{equation}
\kappa=\sqrt{4\Omega^2 + R\dfrac{d\Omega^2}{dR}},
\end{equation}
where $R$ is the radius and $\Omega$ is the angular frequency $\Omega = v_\mathrm{circ}/R$, where $v_\mathrm{circ}$ is the circular velocity. A higher velocity dispersion means higher gas pressure stabilizing the disc against collapse. Here, we calculate $\kappa$ from our best-fitting arctangent models. $\kappa$ thus depends on the potential, but does not include the effects of any streaming or non-circular motions. 

Theoretically, the disc is expected to be unstable to perturbations if $Q\lesssim1$. However, star-forming galaxies have been found to have $Q$ values above unity \citep[e.g.][]{Leroy2008, Romeo2017}. Higher threshold values have therefore also been used in order to account for asymmetric perturbations not captured in the simple model. Such perturbations are likely present in our disturbed galaxies as well and we therefore caution the reader that $Q$ by itself can probably not determine the stability of the disturbed gas discs. 

Another problem is that the observed gas discs are nearly edge-on and our kinematic modelling therefore suffers from degeneracies due to non-circular rotation and beam smearing effects, although our modelling procedure attempts to take this into account. The derived velocity dispersion found is thus potentially a combination of the true velocity dispersion and the radial variation in velocities and may be spuriously high \citep{Barth2016}. We therefore consider our values upper limits to the true velocity dispersion of the gas.

\begin{table*}
\begin{center}
\caption{\label{tab:Q} \small Resulting parameter values from the kinematic modelling: source identifier, inclination, gas velocity dispersion (upper limit), Toomre stability parameter based on modelled $\sigma_\mathrm{gas}$ and radially averaged, based on modelled $\sigma_\mathrm{gas}$ and gas surface density-weighted, based on $\sigma_\mathrm{gas} = 8$~km~s$^{-1}$ and radially averaged and based on $\sigma_\mathrm{gas} = 8$~km~s$^{-1}$ and surface density-weighted, inclination of the gas disc, and estimate for the merger age equal to three times the dynamical time at the radius out to which the gas is undisturbed.}
\vspace{-3mm}
\begin{tabular}[t]{lclllccr}
\hline
\hline \\[-3mm]
source identifier & inclination & $\sigma_\mathrm{gas}$ & $\langle Q\rangle_\mathrm{R}^{\raisebox{1.5pt}{$\scriptstyle{\rm max}$}}$ & $\langle Q\rangle_{\Sigma_\mathrm{gas}}^{\raisebox{1.5pt}{$\scriptstyle{\rm max}$}}$ & $\langle Q\rangle_\mathrm{R}^\mathrm{8~km/s}$ & $\langle Q\rangle_{\Sigma_\mathrm{gas}}^\mathrm{8~km/s}$ & $t_\mathrm{merge}$ \\
&  (\textdegree) & (km~s$^{-1}$) & & & & & (Myr) \\
\hline \\[-4mm]                                                                                                                                       
2MASXJ09033081--0106127 & 85 & $<40$ & $<16.3$ & $<12.6$ & 3.3 & 2.5 & 109 \\
2MASXJ14161186+0152048  & 76 & $<26$ & $<9.1$ & $<7.3$ & 2.8 & 2.3 & 149 \\
NGC4797                               & 69 & $<16$ & $<19.6$ & $<22.7$ & 9.8 & 11.4 & 39 \\
2MASXJ13010083+2701312  & 78 & $<24$ & $<3.3$ & $<2.1$ & 1.1 & 0.7 & 171 \\
2MASXJ13333299+2616190  & 83 & $<6$ & $<3.1$ & $<3.8$ & 4.1 & 5.0 & $>99$ \\
CGCG013--092                     & 78 & $<16$ & $<5.6$ & $<4.3$ & 2.8 & 2.2 & 128 \\[-1mm]
\hline
\hline                                                                                                                                                
\end{tabular}
\end{center}
\end{table*}   
The results from our kinematic modelling are listed in Table~\ref{tab:Q}. The second and third column give the best fit value for the galaxies' inclinations (with errors $<3$\textdegree) and velocity dispersions. The next four columns show the derived values for $Q$ based either on the best-fit velocity dispersions (fourth and fifth columns) or on a canonical $\sigma_\mathrm{gas}=8$~km~s$^{-1}$ (sixth and seventh columns; \citealt{Caldu2016}). The values given include both the radial averages (fourth and sixth columns) and the surface density-weighted averages (fifth and seventh columns). The derived upper limits for the velocity dispersions are generally high, except for 2MASXJ1333. Even when assuming a canonical value of $\sigma_\mathrm{gas}=8$~km~s$^{-1}$, we find that $Q>1$ and therefore theoretically stable against gravitational collapse, with the exception of 2MASXJ1301. However, these values are similar to those found in local spiral galaxies and therefore do not provide much power to discriminate between galaxies with normal and suppressed SFEs. 

The last column in Table~\ref{tab:Q} lists the estimated merger age, $t_\mathrm{merge}$, assumed to be equal to five times the dynamical time at the transition between the undisturbed disc (based on our modelling) and disturbed gas structures (measured at 3 times the RMS level). Because the gas inside this radius appears to be relaxed, whereas the gas outside this radius is clearly not settled yet, this time can serve as a proxy for when the merger happened. This rough estimate is based on theoretical studies of the relaxation of misaligned gas discs in the potential of elliptical galaxies that found that the relaxation process typically takes a few dynamical times ($t_\mathrm{dyn}$; \citealt{Tohline1982, Lake1983}). Specifically, \citet{Lake1983} found that the relaxation time, i.e. the time it takes a misaligned disc to settle into the plane, was approximately $t_\mathrm{dyn}/\epsilon$, where $\epsilon$ is the eccentricity of the potential. For a typical lenticular galaxy $\epsilon\approx0.2$ \citep{Mendez2008}, so we therefore assume $t_\mathrm{merge}\approx5t_\mathrm{dyn}$. The true value of $t_\mathrm{merge}$ could also be higher if it takes more than five dynamical times for the gas to settle \citep{Voort2015b}.  Our merger ages are lower than the ages of the last starburst derived by \citet{Kaviraj2012} from broadband optical colours. Galaxy 2MASXJ1333 has a lower limit for $t_\mathrm{merge}$, because we found no disturbed gas structures in this object.

\subsection{Kennicutt-Schmidt relation} \label{sec:KS}

\begin{figure}
\center
\includegraphics[scale=0.52]{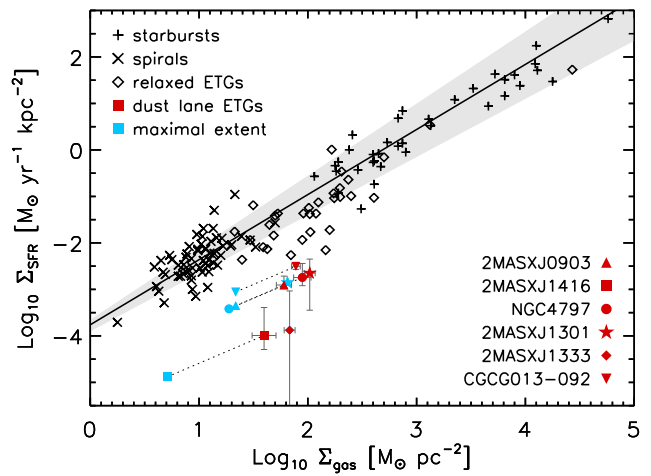}
\caption {\label{fig:KS} Comparison between the molecular gas surface density ($\Sigma_\mathrm{gas}$) and SFR surface density ($\Sigma_\mathrm{SFR}$). The black, solid line and grey shaded region show the average relation for star-forming galaxies and its $1\sigma$ scatter \citep{Kennicutt1998}. Local spiral galaxies (black crosses) and starburst galaxies (black plusses) fall on this relation, whereas relaxed ETGs (open black diamonds) are slightly offset, on average \citep{Davis2014}. Our six dust lane ETGs are shown as red symbols, with $1\sigma$ error bars, for which we take the area of the undisturbed gas that can be modeled as a disc. The cyan symbols (connected to the corresponding red symbols by a dotted line) instead take into account the maximum extent of the disturbed features by assuming this gas is also located in a symmetric disc (note that galaxy 2MASXJ1333 shows no disturbed features). This results in the maximum possible surface area of the gas (both disturbed and undisturbed). This overestimate of the surface area moves the galaxies slightly closer to the $\Sigma_\mathrm{SFR}-\Sigma_\mathrm{H_2}$ relation, but not enough to affect our conclusions. Our sample of ETGs that have undergone a recent minor merger have molecular gas surface densities similar to spiral galaxies and relaxed ETGs, but much lower SFR surface densities, falling on overage 2 orders of magnitude below the Kennicutt-Schmidt relation. }
\end{figure} 
Because we spatially resolve the gas disc in all our dust lane galaxies that recently experienced a gas-rich merger, we can compare them to the observed Kennicutt-Schmidt relation \citep{Kennicutt1998}. Figure~\ref{fig:KS} shows the relation between SFR surface density, $\Sigma_\mathrm{SFR}$, and molecular gas surface density, $\Sigma_\mathrm{gas}$ for our six SFE suppressed ETGs (red symbols with $1\sigma$ error bars) compared to galaxies in the literature. This is calculated by dividing the SFR and $M_\mathrm{H_2}$ by the area of the molecular gas disc. The black, solid line and grey shaded region show the average relation for star-forming galaxies and its $1\sigma$ uncertainty, based on local spiral galaxies (black crosses) and starburst galaxies (black plusses) from \citet{Kennicutt1998}. Relaxed ETGs (black diamonds) from the ATLAS$^\mathrm{3D}$ sample fall below the Kennicutt-Schmidt relation, on average \citep{Davis2014}. 

To estimate the surface area of the molecular gas in our targets, we use our kinematic models to estimate the extent of the relaxed gas disc (measured at 3 times the RMS level, which is approximately 7~M$_\odot$pc$^{-2}$) and assume axisymmetry. The resulting surface densities are shown as the red, filled symbols. The $1\sigma$ errors are based on the errors in the SFR and $M_\mathrm{H_2}$ measurements. Since there are disturbed gas features outside the disc, the surface area may in fact be somewhat larger. We estimate the maximal extent by assuming all the detected gas is located in an axisymmetric disc and show the resulting surface densities as cyan, filled symbols (except for galaxy 2MASXJ1333, because its gas disc is not disturbed). Error bars have been omitted, but are the same as for the corresponding undisturbed measurements (i.e.\ the red symbols). The true values will lie in between our two estimates (connected by dotted lines), but will likely be much closer to the one which excludes the disturbed gas (i.e.\ the red symbols). 

Note that \citet{Kennicutt1998} included the atomic gas as well as the molecular gas within the star-forming region, whereas we (and \citealt{Davis2014}) used only the total molecular gas mass within the molecular gas region (which is likely similar to the star-forming region). H\,\textsc{i} has been detected in the majority of our dust lane ETGs \citep{Davis2015}, but this emission was unresolved. Adding H\,\textsc{i} to our $\Sigma_\mathrm{gas}$ determination, would move our points to the right in Figure~\ref{fig:KS} (by $0.2-0.5$~dex), only exacerbating the disagreement with the star-forming galaxies. 

The $\Sigma_\mathrm{gas}$ and $\Sigma_\mathrm{SFR}$ estimate based on the undisturbed area as well as the one based on the disturbed area show that these dust lane ETGs have much lower star formation surface densities than spiral galaxies and relaxed ETGs at fixed $\Sigma_\mathrm{gas}$. Because our galaxies have molecular gas densities similar to those in spirals and relaxed ETGs, the suppression is not caused by low gas densities at least in an average sense. It remains possible that the dense molecular gas, traced by molecules such as HCN and closely related to the SFR \citep{Gao2004}, has lower surface densities in our objects than in relaxed systems. Also if this is the case, the question remains which physical process is responsible for this.

\section{Discussion and conclusions} \label{sec:concl}

We studied the morphology and kinematics of cold, molecular gas in six dust lane ETGs. These objects were selected to have very low SFEs. These systems were observed previously and are known to exhibit mild disturbances in their stellar distribution, suggesting that they underwent minor mergers in the recent past. It is possible that star formation is suppressed due to the fact that the gas brought in by the minor merger is not dynamically relaxed. If the gas is still streaming in to settle into a disc, the high velocity may have a stabilizing effect on the gas, preventing it from collapsing. We obtained spatially resolved molecular gas measurements in these galaxies with ALMA to test this hypothesis. 

We detect edge-on gas discs in all our six targets, as expected. The majority of these show disturbed features, such as detached gas clouds, warps, and other asymmetries, both morphologically and kinematically. Warps are expected when the gas, which was brought in via a minor merger with its angular momentum misaligned from the stars in the host galaxy, is still settling into the stellar potential \citep[e.g.][]{Voort2015b}. Even though 5 out of 6 galaxies show some assymetries indicating the gas is not fully relaxed, the majority of the molecular gas is located in the central disc, which exhibits regular rotation. 

To better understand the kinematic behaviour of the gas, we fit the data cube with a kinematic model using the KinMS modelling code coupled to an MCMC code. The resulting velocity dispersions of the molecular gas are relatively high ($16-40$~km~s$^{-1}$), except in galaxy 2MASXJ1333 (6~km~s$^{-1}$; which also shows no disturbed features). This, in principle, supports our claim that the gas discs are not dynamically relaxed. However, due to the fact that we are viewing the discs edge-on, we are unsure how much these velocity dispersions are biased because of projection effects (caused by non-circular motions). Future observations of more face-on galaxies with low SFEs (with inclinations below 60\textdegree) will enable improved modelling, especially when paired with high-resolution high resolution optical or near-infrared imaging to better constrain the galaxy rotation curve. 

We believe our objects have only recently acquired their molecular gas and are therefore in a special phase of their evolution in which the gas is not yet in dynamical equilibrium with the other galaxy components. Based on the observed disturbances in the gas distribution, we derive (minor) merger ages between 39~and 171~Myr. The gas is likely still flowing towards the centre, which potentially affects their stability against fragmentation and star formation. This enhanced stability could be provided by inflowing motions \citep[e.g.][]{Meidt2013} or by shear \citep{Seigar2005} or by excess (weak) shocks or (magnetohydrodynamic) turbulence in the molecular gas \citep[e.g.][]{Cluver2010, Padoan2011}. Other observations of star formation in tidal tails or ram pressure stripped tails have also revealed low SFEs \citep[e.g.][]{Knierman2013, Jachym2014}. Certain interacting galaxies are observed to have a large fraction of warm molecular gas, likely due to shocks and turbulence induced by gas accretion, and to show suppressed SFEs \citep{Alatalo2014, Appleton2014}. These are consistent with our interpretation that the suppression of the SFE is due to dynamical effects. 

The SFRs measured are sensitive to star formation in the past $\approx100$~Myr. This timescales is similar to our estimated merger ages. However, the sensitivity increases towards more recent star formation and is most sensitive to stars with ages $\lesssim5-10$~Myr. It is possible that the star formation rate in our objects is increasing with time, which would mean that the instantaneous SFR is higher than the 100~Myr-averaged SFR, in which case we would be underestimating the SFE in our recently merged galaxies. However, for this to negate the $1-2$ orders of magnitude suppression of the SFE, the SFR would have to rise by a similarly large factor over the past $5-10$~Myr for all of our objects, which we consider unlikely. 

Although the lack of synchrotron emission at 3~mm argues against any active supermassive black holes in our galaxy sample, AGN feedback could eject cold molecular gas and suppress star formation. However, \citet{Rosario2017} find normal gas fractions and SFEs in the centres of nearby Seyfert galaxies. Additionally, outflows are generally detected away from the disc and emanating from the centre, whereas the kinematically disturbed features we detected are located in the plane of the disc. Furthermore, since the molecular gas is very extended, nuclear feedback is unlikely to efficiently couple to it. We therefore disfavour the explanation that AGN feedback (or supernova feedback) could be responsible for the abnormally low SFEs. 

ETGs that have recently undergone a minor merger are not the only objects that feature severely suppressed SFEs. Similarly strong suppression of the SFE has been found in post-starburst galaxies \citep{Kohno2002, French2015, Suess2017}. These are very different objects, because their star formation rates are lower than in the recent past, since they just experienced a starburst, whereas star formation rates in our ETGs are higher than before, since the star-forming gas was recently brought in by a gas-rich merger. The physical reason behind the SFE suppression may be different in these different types of galaxies, but spatially resolved observations of the molecular gas content in post-starbursts may help answer this question. 

In summary, we obtained resolved observations of CO emission in six dust lane ETGs with known low SFEs. We find clear morphological and kinematic disturbances, which indicate that the suppression of star formation could indeed be due to the gas motion stabilizing the molecular gas against collapse. Gas-rich mergers with gas-rich hosts are known to result in starburst events with elevated SFEs, due to the compression of the galaxies' ISM. The opposite effect is found here, due gas-poor nature of the hosts studied in this work and therefore the absence of strong shocks. Future studies of these effects in hydrodynamical simulations will be useful to understand the evolution of the SFE during major and minor, gas-rich and gas-poor mergers.

\section*{Acknowledgements}

We would like to thank the referee for their comments that helped clarify the manuscript. We would also like to thank Sugata Kaviraj for helpful suggestions. 
FvdV is supported by the Klaus Tschira Foundation. 
TAD acknowledges support from a Science and Technology Facilities Council Ernest Rutherford Fellowship. 
SSS thanks the Australian Research Council for an Early Career Fellowship, DE130101399.
Parts of this research were conducted by the Australian Research Council Centre of Excellence for All Sky Astrophysics in 3 Dimensions (ASTRO 3D), through project number CE170100013.
YST is supported by the Carnegie-Princeton Fellowship and the Martin A. and Helen Chooljian Membership from the Institute for Advanced Study in Princeton.
This paper makes use of the following ALMA data: ADS/JAO.ALMA\#2015.1.00320.S and we thank all those involved in the proposal. ALMA is a partnership of ESO (representing its member states), NSF (USA) and NINS (Japan), together with NRC (Canada), MOST and ASIAA (Taiwan) and KASI (Republic of Korea), in cooperation with the Republic of Chile. The Joint ALMA Observatory is operated by ESO, AUI/NRAO and NAOJ. This research has made use of the NASA/IPAC Extragalactic Database (NED) which is operated by the Jet Propulsion Laboratory, California Institute of Technology, under contract with NASA.

\bibliographystyle{mnras}
\bibliography{sfsuppression}

\bsp

\label{lastpage}

\end{document}